\documentstyle[12pt]{article}

\begin{document}

\begin{titlepage}

\begin{flushright}
{CERN-TH/2003-055}
\end{flushright}

\vspace*{0.2cm}
\begin{center}

\large {\textbf{Challenges and Obstacles for a\\
Bouncing Universe in Brane Models}}\\
\vspace*{5mm}
\normalsize

{\textbf{P. Kanti and K. Tamvakis\footnote{On leave from the University
of Ioannina, Ioannina, Greece}}}

\smallskip
\medskip
{\textit{CERN, Geneva, Switzerland}}

\smallskip
\end{center}
\vskip0.6in

\centerline{\large\textbf {Abstract}}

A Brane evolving in the background of a charged AdS black-hole displays in general a 
{\textit{bouncing}} behaviour with a smooth transition from a contracting to 
an expanding phase. We examine in detail the conditions and consequences of
this behaviour in various cases. For a cosmological-constant-dominated Brane, we
obtain a singularity-free, inflationary era which is shown to be compatible only
with an intermediate-scale fundamental Planck mass. For a radiation-dominated Brane,
the bouncing behaviour can occur only for background-charge
values exceeding those allowed for non-extremal black holes. For a 
matter-dominated Brane, the black-hole mass affects the proper volume or
the expansion rate of the Brane. We also consider the Brane evolving  in an
asymmetric background of two distinct charged AdS
black hole spacetimes being bounded by the Brane and find that, in the case
of an empty critical Brane, bouncing behaviour occurs only if the black-hole
mass difference is smaller than a certain value. The effects of a Brane
curvature term on the bounce at early and late times are also investigated.

\end{titlepage}

\section{Introduction}

The idea of realizing our universe as a defect \cite{VAR} in a higher-dimensional
spacetime has received a lot of attention in the recent years after the
introduction of D-Branes \cite{Pol},
i.e. membranes on which the fundamental string fields satisfy Dirichlet
boundary conditions.  Motivated by String/M-theory \cite{HW} and the AdS/CFT
correspondence \cite{MGW}, Brane models have revealed new possibilities for the
resolution of the hierarchy problem of particle physics
\cite{A} \cite{NAH}\cite{ANAH}\cite{RS1}. The D-Brane is assigned an intrinsic
energy density and pressure arising both from an underlying Brane tension and
from ordinary $(3+1)$-dimensional matter trapped on it by stringy effects.
Gravitons, on the other hand, propagate into the higher-dimensional bulk.
Nevertheless, as it turns out in the Randall-Sundrum model (RS)
\cite{RS2}\cite{GOG}, virtual gravitons are localized near the Brane due 
to the curvature of the higher-dimensional bulk. In this model, our spacetime
is embedded in a higher dimensional space with an extra highly curved but
infinite fifth dimension. The localization distance of gravity is proportional
to the characteristic length defined by the cosmological constant of the 
Anti de Sitter (AdS)
bulk space. While the Poincare-invariant RS solution requires a fine-tuning 
of the Brane tension, non-Poincare-invariant solutions are also possible. A
four-dimensional Friedman-Robertson-Walker (FRW) universe can arise on a Brane
embedded in an AdS bulk~\cite{CS} or an AdS-black-hole bulk
spacetime~\cite{KR}\cite{KK}. However, in both cases, the presence of extra terms, 
remnants from the higher-dimensional theory, may lead to modifications 
in the evolution of the Brane at small scales.

The Brane-World framework that will be followed in this paper consists of our
physical universe being regarded as a (3+1)-dimensional hypersurface embedded
in a (4+1)-dimensional AdS bulk. The recent observational evidence of
cosmological acceleration motivates the consideration not only of a
{\textit{critical}} Brane of a vanishing four-dimensional cosmological constant
but also of a {\textit{non-critical}} Brane as well. The bulk space background
will be taken to be that of a (4+1)-dimensional AdS black hole 
\cite{KR}\cite{PETK}\cite{MED1} with charge~\cite{X}. Recent
investigations
\cite{PEL}\cite{MED2} seem to indicate that, due to the non-vanishing
charge, a {\textit{bouncing}} universe could, in principle, arise, i.e. a
universe that bounces from a contracting phase to an expanding one without
encountering a singularity (see also \cite{KK} and \cite{Coule}-\cite{Biswas}; 
for earlier examples of singularity-free solutions in the framework
of superstring theory, see \cite{NS}). 
Reference \cite{PEL} considers a semi-realistic radiation-dominated Brane,
while Ref. \cite{MED2} studies a generally non-critical but empty Brane. 

In the present article, we extend existing studies considering a Brane evolving
in a charged AdS black hole background. After presenting the theoretical
framework of our analysis and briefly reviewing the derivation of the
Friedmann equation on the Brane, we reconsider the evolution of both a critical
and non-critical Brane. In the former case, we reconfirm the occurrence of
a bounce at small scales that renders the solutions free from both past and
future singularities. In the latter case, the singularity-free, early regime is
followed by an asymptotically expanding de Sitter epoch, the sequence of which
successfully models an early, inflationary period. We find that the asymptotic
Hubble parameter for the expansion on the Brane is bounded from above by the
black-hole mass and that this model is compatible only with
intermediate-scale gravitational theories, i.e. with $M_5 > 10^{-5} M_P$.
We then proceed to study the evolution of a radiation-dominated Brane and to
derive the exact solution for the scale factor of the four-dimensional subspace,
which is indeed characterized by a non-vanishing minimum value. A careful
examination, however, reveals that the bouncing behaviour for a 
radiation-dominated Brane occurs for background-charge values exceeding those
allowed for non-extremal black holes. Finally, in the case of a Brane filled
with a matter energy-density, the presence of the charge bulk parameter ensures
once again the avoidance of the future singularity in the case of a closed
universe. The presence of the black-hole mass, that survives in the
Friedmann equation at large scales during the same period, also affects
the evolution on the Brane: the main implication for a closed universe is
the increase in its proper volume while, for a flat and open universe, this
term increases or decreases, respectively, the rate of expansion.

We then proceed to consider a Brane evolving in an asymmetric background of two
distinct charged AdS black-hole spacetimes being bounded by the Brane. In the
case of an empty critical Brane, we find that, for the occurrence of a bounce,
the asymmetry in the black-hole mass parameters has to be smaller than a certain
value, in contrast to the symmetric case where bouncing behaviour occurs
generically. For a radiation or matter-dominated Brane, the effect of the
asymmetry is irrelevant both at small and large scales. We finally assume
the presence of an intrinsic boundary curvature term in the action which is
expected to modify the evolution of the universe only at large scales. We
show that, indeed, the value of the scale factor at the bouncing point does
not depend on this term, even for large values of the associated parameter
that determines the magnitude of this term in the action, and that the only
effects coming from the boundary curvature term are relevant in the large
scale factor regime.

\section{The $\bf{(4+1)}$-dimensional Charged Black Hole Background}

We shall consider the following $(4+1)$-dimensional gravitational theory
described by the action 
\begin{equation}
{\cal{S}}=\frac{1}{16\pi G}\int_M d^5x\sqrt{-g}\left({\cal R}_5+
\frac{12}{\ell^2}- F_{MN} F^{MN}\right) +
\frac{1}{8\pi G}\int_{\partial M} d^4 x \sqrt{-\gamma}\,{\cal{K}}\,,
\end{equation}
where ${\cal R}_5$ denotes the scalar curvature of the 5-dimensional spacetime,
$\ell$ is the AdS curvature length related to the {\textit{bulk cosmological
constant}} through
$\Lambda_5=-6/\ell^2$, and $F_{MN}$ stands for the field strength of
a bulk gauge field. The bulk space $M$ consists in general of two different
regions separated by the hypersurface $\partial M$ signifying the Brane, the
simplest choice being two regions related by a ${\cal{Z}}_2$-symmetry.
${\cal{K}}$ is the trace of the {\textit{extrinsic curvature}} on
$\partial M$ defined as ${\cal{K}}_{MN}=\nabla_M\eta_N$ in terms of the
unit normal on it. Finally, $\gamma_{\mu\nu}$ is the induced metric on the
boundary, and  $G$ the five-dimensional Newton constant. 

In addition to the above, we assume a term $\int d^4x\sqrt{-\gamma}\,{\cal L}$
from which a conserved four-dimensional energy-momentum tensor $T_{\mu\nu}$
arises, satisfying the {\textit{Israel Junction Conditions}} 
\begin{equation}
\Delta {\cal{K}}_{\mu\nu}\equiv {\cal{K}}_{\mu\nu}^{(+)}-
{\cal{K}}_{\mu\nu}^{(-)}=-8\pi G\left(T_{\mu\nu}-\frac{1}{3}
T_{\lambda}^{\lambda}\gamma_{\mu\nu}\right)\,.
\end{equation}

Einstein's equations in the bulk are satisfied by the {\textit{AdS charged
black hole}} background metric
\begin{equation}
ds_5^2=-f(r)dt^2+f^{-1}(r)dr^2+r^2d\Omega_{3,k}^2
\end{equation}
with
\begin{equation}f(r)\equiv \frac{r^2}{\ell^2}+k-\frac{\mu}{r^2}+
\frac{q^2}{r^4}\,.
\end{equation} 
In the above, $d\Omega_{3,k}^2$ stands for a $3D$ spatial geometry with the
topology of a plane ($k=0$), a sphere ($k=1$) or a hyperboloid ($k=-1$). The
parameters  appearing in the metric function $f(r)$ are related to the ADM mass
and charge parameters of the black hole through 
\footnote{The charge $Q$ is associated with an Abelian gauge field defined in
the Bulk and has nothing to do with the usual electric charge carried by Brane
matter. All Standard Model fields are assumed to be strictly localized on the
Brane. The two black-hole bulk spacetimes, whose common boundary
is the 3-brane, are characterized by charges of equal value but opposite
signs, an assumption which is consistent with  the ${\cal{Z}}_2$-symmetry.
In this way, the lines of the abelian bulk field start from the positive
charge and end at the negative one, extending continuously over the brane,
and the abelian flux is conserved without having to introduce additional
charges on the brane. By using AdS/CFT correspondence, one can derive
the form of the potential at the location of the brane
\cite{PEL}\cite{Myung}\cite{CAI}\cite{E}
which reads $\Phi=\ell \phi/r$, where $\phi$ is the 5-dimensional
potential given by $\phi=3 \omega_4 Q/8 r^2$.}
\begin{equation}
\mu\equiv \omega_4 M\,\,\,,\,\,\,\,\,\,\,\,\,q^2\equiv 3\omega_4^2Q^2/16\,,
\end{equation}
with $\omega_4=16\pi G/3\Omega_3=8G/3\pi$.

The black hole possesses in general two horizons the position of which is
determined by the solution of the cubic equation $f(r)=0$. It will be useful,
at this point, to introduce the dimensionless parameters and variables
$\overline{\mu}\equiv \mu/\ell^2$, $\overline{q}^2\equiv q^2/\ell^4$ and
$y\equiv r^2/\ell^2$. Then, the case of {\textit{two distinct horizons}}
corresponds to values of the charge
\begin{equation}
\overline{q}^2<\overline{q}_+^2\,,\end{equation}
where
\begin{equation}
\overline{q}_{\pm}^2\equiv -\frac{k}{3}\left(\overline{\mu}+
\frac{2}{9}k^2\right)\pm \frac{2}{3\sqrt{3}}\left(\overline{\mu}+
\frac{k^2}{3}\right)^{3/2}\,,
\end{equation}
Note that always $\overline{q}_+^2>0$ and $\overline{q}_-^2<0$. The two
horizons correspond to the two positive solutions of a cubic equation,
namely,
\begin{equation}
y_{out}=-\frac{k}{3}+2\left(\frac{\overline{\mu}}{3}+
\frac{k^2}{9}\right)^{1/2}\cos(\phi-\pi/3)\\,
\end{equation}
\begin{equation}
y_{in}=-\frac{k}{3}+2\left(\frac{\overline{\mu}}{3}+
\frac{k^2}{9}\right)^{1/2}\sin(\pi/6-\phi)\,,
\end{equation}
where we have introduced 
\begin{equation}
\phi\equiv\frac{1}{3}\tan^{-1}\left(\frac{2}
{\sqrt{\frac{q^2-q_-^2}{q_+^2-q^2}}-
\sqrt{\frac{q_+^2-q^2}{q^2-q_-^2}}}\right)\,.
\end{equation}

For values of the charge larger than the limiting charge, namely for
$q^2>q_+^2$, two of the roots of the horizon equation are complex and there
is only one horizon. Thus, in this case we have an {\textit{extremal black
hole}}. The stability status of extremal black holes is still an open
question~\cite{E} and, perhaps, they should be avoided as a background.

\section{Brane-World in a Charged Black Hole Background}

Following the steps of Ref. \cite{KR} and introducing a spherically symmetric
3-Brane at the position $r=R$, we obtain from the Israel Junction Conditions
the following {\textit{Friedmann Equation}} on the Brane
\begin{equation}
\frac{\left(f(R)+\dot{R}^2\right)^{1/2}}{R}=\frac{4\pi G}{3}\,(\rho+\sigma)\,,
\label{FRW-1}
\end{equation}
where the dot denotes the derivative with respect to the proper time $\tau$
on the Brane. In addition, $\rho$ is the matter energy-density on the Brane and
$\sigma$ the Brane tension. The last two arise from the Brane energy-momentum
tensor $T_0^0=-(\rho+\sigma)$, $T_j^i=\delta_j^ip$, conserved through the
equation
\begin{equation}
\nabla_{\mu}T_{\nu}^{\mu}=0\Longrightarrow \dot{\rho}+
3\frac{\dot{R}}{R}\,(\rho+p)=0\,.
\end{equation}

Note that this equation is derived only in the case of 
${\cal{Z}}_2$-symmetry~\footnote{The radius $r$ is decreasing on both sides of
the Brane.}. In an asymmetric situation, we have the more general equation
\begin{equation}
\frac{\left(f_+(R)+\dot{R}^2\right)^{1/2}}{R}+\frac{\left(f_-(R)+
\dot{R}^2\right)^{1/2}}{R}=\frac{8\pi G}{3}\,(\rho+\sigma)\,.
\label{asymmetric}
\end{equation}
The metric functions $f_{\pm}(r)$ can differ in the vacuum parameters
$\mu_{\pm}$ and $\ell_{\pm}$. 

The four-dimensional metric on the Brane corresponds to a FRW-universe,
$R(\tau)$ being the scale factor. It is
\begin{equation}ds_4^2=-d\tau^2+R^2(\tau)\,d\Omega_{3,k}^2\,.\end{equation}
Taking the square of the Friedmann Equation (\ref{FRW-1}), we obtain the
more conventional form
\begin{equation}H^2\equiv \left(\frac{\dot{R}}{R}\right)^2=-\frac{k}{R^2}+
\frac{\mu}{R^4}-\frac{q^2}{R^6}+\frac{8\pi G_4}{3}\,\Lambda_4+
\left(\frac{4\pi G}{3}\right)^2\left(\rho^2+2\sigma \rho\right)\,.
\label{FRW-2}
\end{equation}
In the above, we have defined the {\textit{four-dimensional cosmological
constant}} as\,\footnote{The four-dimensional Newton constant can be read
off from the linear energy-density term to be
$\frac{8\pi G_4}{3}\equiv 2\sigma\left(\frac{4\pi G}{3}\right)^2$.}
\begin{equation}
\frac{8\pi G_4}{3}\,\Lambda_4 \equiv \left(\frac{4\pi G\sigma}{3}\right)^2-
\frac{1}{\ell^2}\geq 0\,.\end{equation}
The generic case is that of a de Sitter Brane. The case $\Lambda_4=0$ of a
{\textit{Critical Brane}} is achieved through the well known fine-tuning
between bulk ($G$, $\ell$) and Brane ($\sigma$) parameters of the
Randall-Sundrum model. 

The above Friedmann Equation of the Brane features a {\textit{dark energy}}
term $\mu/R^4$ that has the same scale dependence as the standard radiation
term\,\footnote{The equation of state $p=w\rho$, through the conservation
equation $\dot{\rho}/\rho=-3(1+w)\dot{R}/R$, in the case of radiation
($w=-1/3$), corresponds to $\rho\propto R^{-4}$.}. The term arising from
the presence of the bulk charge corresponds to a {\textit{stiff energy}} 
equation of state ($w=1$) characterized by an exotic negative energy density.

\section{Review of the Evolution of an Empty Brane}

It is instructive to review, and complete, the solutions of the Friedmann
Equation in the case of an empty Brane 
\cite{PETK}\cite{MED1}\cite{PEL}\cite{MED2} before proceeding to study more
realistic cases. For $\rho=0$, Eq.(\ref{FRW-2}) takes the form
\begin{equation} 
\left(\frac{\dot{R}}{R}\right)^2=-\frac{k}{R^2}+\frac{\mu}{R^4}-
\frac{q^2}{R^6}+\frac{8\pi G_4}{3}\,\Lambda_4\,.
\label{critical}
\end{equation}

We will now consider separately the cases of a critical ($\Lambda_4=0$) and
non-critical ($\Lambda_4 \neq 0$) Brane: 

\medskip \noindent
{\sl 4.1 An Empty, Critical Brane}

By assuming an empty and critical Brane ($\Lambda_4=0$) and introducing a new
time variable $d\tau=R (\eta)\,d\eta$, Eq.(\ref{critical}) leads to the
solutions shown in Table 1, for $k=0, \pm1$. The parameter $\epsilon$ is
defined as $\epsilon^2=4 q^2/\mu^2$. All three solutions are characterized
by a minimum radius of contraction beyond which the universe bounces to an
expanding phase. Thus, there is no primordial or future singularity associated
with these cosmologies. Note that the radius at which the bouncing occurs is
always outside of the outer horizon of the black hole, since
$H^2=-f(R)/R^2+\frac{8\pi G}{3}\,\sigma^2=0$ implies
$f(R)=\frac{8\pi G}{3}\,\sigma^2 R^2>0$.

The solution for $k=1$ is periodic 
and is characterized by an infinite number of bounces at the two points
$R_{min}$ and $R_{max}$. This solution does not possess neither Big Bang nor
Big Crunch singularity and it is possible only for $\epsilon^2<1$. This
restriction on the black hole charge ($q^2<\mu^2/4$) is always satisfied
if the black hole of the background has two horizons. Indeed, recalling
the corresponding constraint $\overline{q}^2<\overline{q}_+^2$, we see
that $\overline{q}_+^2$ is always smaller than $\overline{\mu}^2/4$. A
simple numerical analysis shows that 
$\frac{\overline{\mu}^2}{4}+\frac{1}{3}\left(\overline{\mu}+
\frac{2}{9}\right)-\frac{2}{3\sqrt{3}}\left(\overline{\mu}+
\frac{1}{3}\right)^{3/2}$ is always 

\begin{center}
\begin{tabular}{|c||c||c||l|}\hline \hline
\multicolumn{4}{|c|}{{\bf Table 1:} Charged ($Q^2 \neq 0$), Critical 
($\Lambda_4=0$)\,\,Brane}\\ \hline \hline
\,&\,&\,&\,\\

$k=1$ & $R^2=\frac{\mu}{2}\left(1-\sqrt{1-\epsilon^2}\cos 2\eta\right)$ & 
\begin{tabular}{c}
$R_{min}^2= \frac{\mu}{2}\left(1-\sqrt{1-\epsilon^2}\right)$ \\[4mm]
$R_{max}^2=\mu/2$, $\epsilon^2<1$ \end{tabular} &
\begin{tabular}{c} bouncing\\ cyclic \end{tabular} \\[9mm]
\hline
\,&\,&\,&\,\\
$k=0$ & $R^2=\frac{\mu}{4}\left(\epsilon^2+4\eta^2\right)$ & 
$R_{min}^2=\mu \epsilon^2/4$, $R_{max}^2=\infty$ & 
\begin{tabular}{c} bouncing\\ expanding \end{tabular}\\[6mm]
\hline
\,&\,&\,& \,\\
$k=-1$ & $R^2=\frac{\mu}{2}\left(-1+\sqrt{1+\epsilon^2}\cosh 2\eta \right)$ 
& \begin{tabular}{c}
$R_{min}^2=\frac{\mu}{2}\left(\sqrt{1+\epsilon^2}-1\right)$ \\[4mm]
$R_{max}^2=\infty$\end{tabular} 
& \begin{tabular}{c} bouncing\\ expanding \end{tabular}\\[9mm]
\hline \hline
\end{tabular}
\end{center}

\noindent positive for any value 
$0<\overline{\mu}<\infty$. By using dimensionful parameters,
we may write the allowed range of values of the added black hole charge,
for the existence of physically acceptable bouncing universes in
a two-horizon black hole background, as
\begin{equation}
0<Q^2<\frac{16}{3 \omega_4^2}\,\biggl[-\frac{\ell^2}{3}\left(
\omega_4 M+ \frac{2\ell^2}{9}\right)+\frac{2\ell}{3\sqrt{3}}
\left(\omega_4 M+ \frac{\ell^2}{3}\right)^{3/2}\biggr]\,,
\label{k-1}
\end{equation}
for a given ADM black hole mass $M$ and AdS curvature length $\ell$.
Alternatively, for fixed $M$ and $Q^2$, the above constraint may be
interpreted as a lower bound on the AdS curvature $\ell$ or, through
the relation $\sigma=3/4\pi G\ell$ for a critical Brane, as an upper
bound on the tension $\sigma$ of such a Brane, that is introduced in
the aforementioned background. 

For $k=0,-1$, the solutions are characterized by a single bounce that demands
again a non-vanishing value of the black hole charge. The constraint for the
existence of two horizons still needs to be satisfied and reduces to
\begin{equation}
Q^2 < \frac{32 M^2 \ell}{9 \sqrt{3 \omega_4 M}}\,,
\end{equation}
for $k=0$, and to Eq. (\ref{k-1}) with the sign of the first term on the
rhs reversed, for $k=-1$.

\bigskip
\noindent
{\sl 4.2 An Empty, Non-critical Brane}

In the case of non-vanishing four-dimensional cosmological constant
$\Lambda_4$, the Friedmann Equation is modified only for very large values
of the scale factor $R$. The short distance behaviour is dominated by the
mass and charge terms. Thus, the $\Lambda_4\neq 0$ solutions at short
distances are very close to the previously discussed set, while for large
distances they are very close to the solutions of the Friedmann Equation
with a vanishing charge, since at those distances the charge term becomes
irrelevant. The latter set are given in Table 2, where we have defined a
new parameter $\overline{\epsilon}$ through the relation 
$\overline{\epsilon}^2\equiv 4\kappa^2_4 \mu\Lambda_4/3$, with 
$\kappa^2_4=8 \pi G_4$. 

\begin{center}
\begin{tabular}{|c||c||c||l|}\hline \hline
\multicolumn{4}{|c|}{{\bf Table 2:} Neutral ($Q^2 = 0$), Non-critical 
($\Lambda_4 \neq 0$)\,\,Brane}\\ 
\hline \hline

\,&\,&\,&\,\\

$k=1$ & $R^2=\frac{3}{2\kappa^2_4 \Lambda_4}\left\{1+\sqrt{1-
\overline{\epsilon}^2}
\cosh\left( 2 \kappa_4\sqrt{\frac{\Lambda_4}{3}}(\tau-\tau_0)\right)\right\}$ &
$R_{\infty}^2\propto e^{2\kappa_4\sqrt{\frac{\Lambda_4}{3}}\,\tau}$ & as.dS\\

\,&\,&\,&\,\\
\hline
\,&\,&\,&\,\\

$k=0$ & $R^2=\sqrt{\frac{3\mu}{\kappa^2_4 \Lambda_4}}
\sinh\left(2 \kappa_4\sqrt{\frac{\Lambda_4}{3}}(\tau-\tau_0)\right)$ & 
$R_{\infty}^2\propto e^{2\kappa_4\sqrt{\frac{\Lambda_4}{3}}\,\tau}$ 
& as.dS \\
\,&\,&\,&\,\\
\hline
\,&\,&\,&\,\\

$k=-1$ & $R^2=\frac{3}{2\kappa^2_4 \Lambda_4}\left\{-1+
\sqrt{1-\overline{\epsilon}^2}
\cosh\left( 2 \kappa_4\sqrt{\frac{\Lambda_4}{3}}(\tau-\tau_0)\right)\right\}$ & 
$R_{\infty}^2\propto e^{2\kappa_4\sqrt{\frac{\Lambda_4}{3}}\,\tau}$ & as.dS\\
\,&\,&\,&\,\\
\hline \hline
\end{tabular}
\end{center}

Joining together the two sets of solutions presented in these two subsections,
we see that the cosmology of an empty Brane with a non-vanishing cosmological
constant possesses a bouncing point at early times and can have a generic 
expanding behaviour at late times. This can be a plausible scenario for an
early inflationary era in which the cosmological constant stands for, or
includes, an almost constant energy density of a scalar (inflaton) field.
The bounce at early times guarantees the absence of a Big Bang
singularity\,\footnote{Note, however, that the periodic behaviour of the
critical $k=1$ case that describes a cyclic universe is not retained in
the presence of a cosmological constant.}
for all values of $k$ as long as the bounds on the charge parameter $Q$
presented in the previous subsection are respected. At late times, an
additional constraint arises, for $k=\pm 1$, for the validity of the
solutions, namely $\overline{\epsilon}^2<1$. This constraint leads to an
upper bound on the Hubble parameter of the asymptotic expansion on the
Brane in terms of bulk parameters, namely
\begin{equation}
H^2_\infty \equiv \frac{\kappa^2_4}{3}\,\Lambda_4 < \frac{1}{4 \omega_4 M}
=\frac{3 \pi M_5^3}{32 M}\,,
\label{con-H}
\end{equation}
where $M_5$ is the fundamental scale of gravity in five dimensions. If this
period of asymptotic exponential expansion plays the role of standard
inflation, then the vacuum energy density of the Brane must be of order
$\Lambda_4 \sim (10^{16}\,{\rm GeV})^4$ in order to obtain the correct
magnitude of density perturbations. This, in conjunction to Eq. (\ref{con-H}),
leads to
\begin{equation}
\left(\frac{M_5}{M_P}\right)^2 > 10^{-11} \left(\frac{M}{M_5}\right)\,.
\label{strict}
\end{equation}
Assuming that the black hole mass is at least $M \geq 10\,M_5$,
the above constraint puts a lower bound on the value of the five-dimensional
Planck scale, i.e. $M_5 > 10^{-5} M_P$, in agreement with similar bounds
found in the literature for the occurrence of Brane inflation in 
higher-dimensional models \cite{KO}. Alternatively, pushing the scale of
gravity down to the TeV scale leads to a black hole mass which is many
orders of magnitude below the fundamental scale, a result that invalidates
the classical field theory approach used in our analysis.

\section{Radiation Dominated Brane}

Let us now consider the realistic case of a Brane with a non-zero energy
density that obeys a radiation equation of state ($w=1/3$) and has a scale
factor dependence of the form $\rho=\hat{\rho}/R^4$. Going back to the
Friedmann equation (\ref{FRW-2}), and substituting the energy
density, we obtain
\begin{equation}
H^2=\left(\frac{\dot{R}}{R}\right)^2=\frac{8\pi G_4}{3}\,\Lambda_4-
\frac{k}{R^2}+ \left(\mu+\frac{8\pi G_4}{3}\,\hat{\rho}\right)\frac{1}{R^4}
-\frac{q^2}{R^6}+\frac{4\pi G_4}{3\sigma}\,\frac{\hat{\rho}^2}{R^8}\,.
\end{equation}
In what follows, we solve the above equation both for early and late
times and proceed to check the validity of the derived bouncing solution.

\bigskip \noindent
{\sl 5.1 Derivation of the solution}

There are two distinct scale regimes at which different terms dominate. 
For small scale factors, we may neglect the cosmological constant $\Lambda_4$
and the curvature term $k/R^2$.
This is the {\textit{early regime}} that should be responsible for the
existence of a bounce and the avoidance of the primordial singularity.
Introducing the new variables $x\equiv R^2$ and $d\tau=R^2d\overline{\tau}$,
the approximate Friedmann equation can take the form
\begin{equation}\frac{1}{4}\,(x')^2=bx^2-q^2x+a\,,
\label{rad}
\end{equation}
where we have defined
\begin{equation}
a \equiv \frac{4\pi G_4\hat{\rho}^2}{3\sigma}\,\,\,,\,\,\,\,\,\,
b\equiv \mu+\frac{8\pi G_4}{3}\,\hat{\rho}=
(1+\lambda)\,\frac{8\pi G_4}{3}\,\hat{\rho}\end{equation}
and the prime denotes differentiation with respect to the {\textit{new time}}
$\overline{\tau}$. We have also introduced, for later use, a new
dimensionless parameter $\lambda$ defined through the relation
$\mu=\lambda\,(\frac{8\pi G_4}{3}\,\hat{\rho})$. 
By setting $x'=0$ and demanding the existence of a bounce, the following
condition on the {\textit{minimum value}} of the charge parameter emerges
\begin{equation}
q^4\geq 4ab=\frac{2\hat{\rho}}{\sigma}\,(1+\lambda)
\left(\frac{8\pi G_4}{3}\hat{\rho}\right)^2\,.\label{minimum}\end{equation}
On the other hand, the solution of Eq. (\ref{rad}) has the form
\begin{equation}
R^2=\frac{q^2}{2b}+\sqrt{\Delta}\cosh\left(2\sqrt{b}\overline{\tau}\right)\,,
\label{sol-r}\end{equation}
where $\Delta=-\frac{a}{b}+\left(\frac{q^2}{2b}\right)^2>0$. The time variables
are related through
\begin{equation}
\tau=\frac{q^2}{2b}\,\overline{\tau}+\sqrt{\frac{\Delta}{4b}}
\sinh\left(2\sqrt{b} \overline{\tau}\right)\,.\end{equation}
It is clear that there is a non-zero minimum value of the scale factor
\begin{equation}R_{min}^2=\frac{q^2}{2b}+\sqrt{\Delta}\,,\label{min}
\end{equation}
where the bounce occurs (an arbitrary integration constant in Eq. (\ref{sol-r})
has been chosen such that the point $\overline\tau=0$ coincides with the
time of the bouncing).

For large scale factors, the charge term, as well as the quadratic
energy-density term, due to their scale factor dependence, are suppressed and
thus can be neglected. The resulting Friedmann equation for this {\textit{late
regime}} is
\begin{equation}H^2 \simeq \frac{8\pi G_4}{3}\,\Lambda_4-\frac{k}{R^2}+
\frac{b}{R^4}\,,
\end{equation}
and coincides in form with the one for a non-critical empty Brane. The 
corresponding solutions therefore can be obtained from Table 2 of section 4
with the replacement $\mu\rightarrow b$ and 
$\overline{\epsilon}^2\rightarrow 4b\kappa^2_4\Lambda_4/3$.
They all describe an asymptotically de Sitter expanding universe. Combining
the derived early and late time solutions, we can successfully model an
early, singularity-free, radiation-dominated epoch that passes smoothly to an
inflationary period for the universe.  In this case, the constraint
$\overline{\epsilon}^2 < 1$, puts an upper bound on the asymptotic Hubble
parameter during inflation, i.e.
\begin{equation}
H^2_\infty \equiv \frac{\kappa^2_4}{3}\,\Lambda_4 < \frac{1}{4}\,
\left(\omega_4 M + \frac{8 \pi}{3}\,\frac{\hat\rho}{M_P^2}\right)^{-1}\,,
\end{equation}
in terms of the black hole mass as well as the energy-density of the
precedented radiation-dominated epoch. Viewing the above inequality as
a constraint on the ratio between the five and four-dimensional Planck scales,
we obtain a bound which is even stricter than the one derived in the case
of an empty, non-critical Brane. For negligible values of the parameter
$\hat \rho$, we recover Eq. (\ref{strict}) and the constraint for intermediate
gravity scale, while for large values of $\hat \rho$, $M_5$ is pushed further
towards $M_P$.

In the case of a vanishing four-dimensional cosmological constant (or
sub-dominant compared to the linear energy-density term), we recover at
late times a standard, radiation-type Friedmann equation that describes a
radiation-dominated epoch well after the end of inflation. The solutions
in that {\textit{late regime}}, for all values of $k$, can be obtained
from the ones presented in Table 1 of section 4 by setting $Q^2=0$ and
$\mu \rightarrow b$. These solutions should duplicate exactly the successful
cosmological predictions for nucleosynthesis. As has been noted in the
literature before~\cite{SARK}\cite{OL}\cite{HAN}\cite{PEL}, this puts a
strong bound on any non-standard contribution to the energy density, and
thus on the black hole mass parameter, that has the same scaling as the
linear radiation term. The dark radiation term generated by it should not
exceed the effect that an additional neutrino species would have on the
value of the $R^{-4}$ coefficient. This amounts to $\mu< 1.13\,G_4 \hat{\rho}$
\,or, equivalently, to $\lambda < 0.14$.

\newpage
{\sl 5.2 Validity of the Bouncing Solution}

The occurrence of a bounce in a radiation-dominated Brane requires, as we saw,
a bulk charge larger than a minimum value that depends on the radiation
energy-density, namely
\begin{equation}q^4\geq \frac{2\hat{\rho}}{\sigma}\,(1+\lambda)
\left(\frac{8\pi G_4}{3}\hat{\rho}\right)^2.\end{equation}
Nevertheless, as we discussed in section 2, the charge of the bulk
background metric cannot increase further than a limiting value $q_+^2$
determined by the mass of the black hole, since beyond that charge the two
horizons merge giving us an extremal black hole the stability of which is
questionable. Since neither the curvature $k$ nor the cosmological constant
are of importance in the regime where the bounce occurs, it is sufficient to
consider this bound in the critical and flat case. It is
\begin{equation}q^2<q_+^2=\ell^4\frac{2}{3\sqrt{3}}\,\overline{\mu}^{3/2}=
\ell\,\frac{2}{3\sqrt{3}}\,\mu^{3/2}\,.\end{equation}
Setting $\mu=\lambda \left(\frac{8\pi G_4}{3}\right)\hat{\rho}$ and 
$\ell^{-2}=\left(\frac{4\pi G_4}{3}\right)\sigma$, we get
\begin{equation}
q^4<\frac{4}{27}\left(\frac{3}{4\pi G_4 \sigma}\right)
\lambda^3\left(\frac{8\pi G_4\hat{\rho}}{3}\right)^3.
\label{con-flat}\end{equation}
The two constraints are compatible if 
\begin{equation}\lambda^3-\frac{27}{4}\,(1+\lambda)=
\Bigl(\lambda+\frac{3}{2}\Bigr)^2\,(\lambda-3)>0\,.\end{equation}
This inequality holds only for $\lambda >3$ and cannot be satisfied for
values as low as $\lambda\sim 0.14$ that follows from the nucleosynthesis
constraint. Thus, unfortunately, the charge value required for the 
occurrence of the bounce corresponds to an extremal black hole background.

An alternative to the two-horizon constraint, that also puts an upper bound
on the value of the charge parameter, can be obtained from the requirement
that the energy-density of the universe at the bouncing point must be larger
than the one at the time of nucleosynthesis, i.e.
$\hat\rho/R_{min}^4 >$ (0.2 MeV)$^4$. This constraint was mentioned
in \cite{PEL} but was not properly addressed as the authors lacked the exact
solution for the scale factor. The value of $R_{min}$ varies as a function
of the parameters $q^2$, $\mu$ and $\hat\rho$ according to Eq. (\ref{min}).
The strongest constraint arises by considering the maximal possible value
of $R_{min}$, and thus the minimal possible value of $\rho$, that
corresponds to large values of $q^2$ and is given by $R^2_{min}\simeq q^2/b$. 
Substituting this value in the expression of the energy-density, we obtain
the constraint
\begin{equation}
q^4 < \frac{(1+\lambda)^2 \hat\rho}{(0.2\,\,{\rm MeV})^4}\,
\left(\frac{8 \pi G_4}{3}\,\hat\rho\right)^2\,.
\label{upper}\end{equation}
The above upper bound on the value of charge parameter replaces 
Eq. (\ref{con-flat}) and is necessary for the validity of the bouncing
solution in an extremal black hole, five-dimensional background. The
requirement, finally, that the quadratic energy-density term is subdominant
compared to the linear one, at the time of nucleosynthesis, leads to
\begin{equation}
\sigma > \frac{1}{2}\,\frac{\hat\rho}{R^4} 
\simeq (0.17\,\,{\rm MeV})^4\,,\end{equation}
a value which is smaller than the one derived in Ref. \cite{PEL}. 


\section{Matter Dominated Brane}

Concluding our study of the evolution of a four-dimensional Brane embedded in
a symmetric, AdS charged black-hole, bulk spacetime, we will now study the case
of a matter equation of state for the energy-density on the Brane. In that case,
we have $\rho=\tilde\rho/R^3$ and the Friedmann equation (\ref{FRW-2}) takes
the form
\begin{equation}
H^2=\left(\frac{\dot R}{R}\right)^2=\frac{8\pi G_4}{3}\,\Lambda_4-
\frac{k}{R^2}+ \frac{\mu}{R^4} + \frac{8\pi G_4}{3}\,\frac{\tilde{\rho}}{R^3}
+\left(\frac{4\pi G_4}{3\sigma}\,\tilde{\rho}^2-q^2\right)\,\frac{1}{R^6}\,.
\end{equation}
Given the relevance of this particular equation of state at late times in the
history of the universe, it would not be meaningful to talk about the existence
or not of an initial singularity. For large values of $R$, the charge as well
as the quadratic energy-density term are subdominant and can be safely dropped.
The $\mu$-term remains and the relevant question is how this term, remnant
of the structure of the 5-dimensional bulk, affects the evolution of the
Brane at the {\textit{late time regime}}. For simplicity, we will consider
again a critical Brane with $\Lambda_4=0$, and solve for the scale factor,
for $k=0,\pm 1$. By using the conformal time coordinate 
$d \tau=R(\eta)\,d\eta$, we obtain the solutions listed in Table 3.

\begin{center}
\begin{tabular}{|c||c||c|}\hline \hline
\multicolumn{3}{|c|}{{\bf Table 3:} Neutral ($Q^2 = 0$), Critical 
($\Lambda_4 = 0$), Matter-Dominated\,\,Brane}\\ 
\hline \hline

\,&\,&\,\\

$k=1$ & $R=\frac{A}{2}\left\{1+\sqrt{1+\frac{4\mu}{A^2}}\,
\sin\left[\eta-\eta_0 + \arctan\left(\frac{2 R_0 -A}{2\sqrt{\mu+A R_0 -R_0^2}}
\right) \right] \right\}$ & 
\begin{tabular}{c} expand.- \\ contract.\end{tabular} \\

\,&\,&\,\\
\hline
\,&\,&\,\\

$k=0$ & $R=R_0+\frac{A}{4}\,(\eta-\eta_0)^2
+\sqrt{\mu + A\,R_0}\,(\eta-\eta_0)$ & 
\begin{tabular}{c} power-low \\ expand. \end{tabular} \\
\,&\,&\,\\
\hline
\,&\,&\,\\

$k=-1$ & $R=\frac{A}{2}\left\{-1+\sqrt{1-\frac{4\mu}{A^2}}\,
\cosh\left[\eta-\eta_0 - \ln\left(\frac{\sqrt{A^2-4 \mu}}
{2\sqrt{\mu+A R_0 +R_0^2} + A +2R_0}\right) \right] \right\}$ & 
\begin{tabular}{c} exponent. \\ expand. \end{tabular} \\
\,&\,&\,\\
\hline \hline
\end{tabular}
\end{center}

In the above, we have defined $A=\bigl(\frac{8 \pi G_4}{3}\bigr)\,\tilde\rho$,
and have denoted with $R_0$ the value of the scale factor at the
beginning of the matter-dominated era, at $\eta=\eta_0$. 

For $k=1$, the matter-dominated Brane first expands and then contracts, in
agreement with the standard Cosmological Model. At the point where $H$, or
equivalently $dR/d\eta$, becomes zero, the universe stops expanding and
then recollapses. This occurs at 
\begin{equation}
R=\frac{A}{2}\,\biggl(1+\sqrt{1+\frac{4\mu}{A^2}}\,\biggr)\,,
\end{equation}
and it clearly corresponds to a larger value of the scale factor compared to
the case where $\mu=0$. The main implication, therefore, of the bulk parameters
on the evolution of the closed, matter-dominated Brane, at large scales, is the
increase of the proper volume of the universe. As the Brane contracts, we will
eventually reach small values of the scale factor for which the charge term
will become dominant again. In that case, the evolution of the Brane would
be governed by the equation
\begin{equation}
R^4\,{\dot R}^2=\mu R^2 - \tilde q^2\,,
\end{equation}
where we have defined 
$\tilde q^2=q^2-\frac{4\pi G_4}{3\sigma}\,\tilde{\rho}^2$, and ignored the
curvature and linear energy-density terms which are now subdominant. Clearly,
the above equation is characterized by the vanishing of $\dot R$ at a finite
value of the scale factor, namely at $R^2_{min}=\tilde q^2/\mu$,
as long as $\tilde q^2>0$, a constraint that puts a lower bound on the charge
parameter. If the constraint for the existence of two horizons
(\ref{con-flat}) had not been violated in the precedented radiation-dominated
era, one could have shown that the two constraints on $q^2$ would have been
indeed compatible, in the matter-dominated era, if 
$\sigma > (4 \pi G_4)^3\,\tilde\rho^4/4\mu^3$. If the alternative upper
bound (\ref{upper}) is used instead, we derive the constraint
\begin{equation}
\sigma > \frac{(0.2\,\,{\rm MeV})^2}{2 (1+\lambda)}\,
\frac{\tilde \rho^2 }{\hat \rho^{3/2}}\,.
\end{equation}

For $k=0$ and -1, the four-dimensional Brane expands forever and no future
singularity is encountered, as expected. In the case of a flat universe,
the $\mu$-term adds a positive contribution to the value of the scale
factor and thus increases the rate of expansion. For an open universe,
however, and for a given time $\eta$, we may easily see that the value
of the scale factor is smaller compared to the one for $\mu=0$, and
therefore the bulk parameter delays the expansion of the universe in
this case. The derived solution is valid as long as 
\begin{equation}
\omega_4 M < \frac{1}{4}\,\left(\frac{8 \pi G_4}{3}\,\tilde \rho\right)^2\,,
\end{equation}
which puts an upper bound on the black hole mass $M$.


\section{Bouncing in an Asymmetric Background}

In this section, we shall consider the possibility of an asymmetric bulk space
consisting of two distinct regions terminating on the Brane. To keep things
simple, we shall consider for both regions a charged AdS black hole geometry 
characterized by the same AdS length $\ell$ and charge $Q$ but with different
black hole masses $M_{\pm}$. We shall denote the two metric functions as 
\begin{equation}f_{\pm}(R)=\frac{R^2}{\ell^2}+k-\frac{\mu_{\pm}}{R^2}+
\frac{q^2}{R^4}\,.\end{equation}
The Friedmann Equation takes the form (\ref{asymmetric}) which can be squared
twice to give 
\begin{equation}
\dot{R}^2+k=\frac{\mu}{R^2}-\frac{q^2}{R^4}+\frac{2R^2}{\ell^2}
\left(\frac{\rho}{\sigma}\right)\left(1+\frac{\rho}{2\sigma}\right)
+\frac{\ell^2}{16 R^6}\frac{(\Delta\mu)^2}{\left(1+\frac{\rho}{\sigma}\right)^2}\,,
\label{asym}\end{equation}
where we have assumed a critical Brane following by making the same fine tuning
as in the symmetric case, namely, 
\begin{equation}\left(\frac{4\pi G \sigma}{3}\right)^2=\frac{1}{\ell^2}=
\frac{4\pi G_4}{3}\,\sigma\,,\end{equation}
and have also defined
\begin{equation}\mu\equiv \frac{1}{2}\,(\mu_++\mu_-)\,, \qquad
(\Delta\mu)^2\equiv (\mu_+-\mu_-)^2\,.\end{equation}
It is straightforward to see that, in the case $\mu_+=\mu_-$, the Friedmann
equation (\ref{FRW-2}) for a symmetric bulk is recovered.

In the case of an empty, critical Brane ($\rho=0$) the above evolution equation
simplifies to
\begin{equation}\left(\frac{\dot{R}}{R}\right)^2+\frac{k}{R^2}=
\frac{\mu}{R^4}-\frac{q^2}{R^6}+\frac{\ell^2(\Delta\mu)^2}{16 R^8}\,.
\label{FRW-asym}
\end{equation}
As a result of the asymmetry, there is a positive term present that opposes the
effects of the charge at small values of the scale factor. In the same 
{\textit{early regime}}, the curvature term can be dropped. Then, the above
equation has exactly the same form as the Friedmann equation in the case of
a critical, radiation-dominated universe with a symmetric bulk, with
$(\Delta\mu)^2$ playing the role of the
quadratic energy-density term, and thus possesses a bouncing solution for large
enough charge. Introducing again  $x=R^2$ and $d\tau=R^2d\overline{\tau}$,
we can bring Eq. (\ref{FRW-asym}) in the form
\begin{equation}\frac{1}{4}(x')^2=\mu x^2-q^2 x+\frac{\ell^2(\Delta\mu)^2}
{16}\end{equation}
from which we obtain the solution
\begin{equation}
R^2=\frac{q^2}{2\mu}\left\{1+\sqrt{1-\frac{\mu\ell^2(\Delta\mu)^2}{4q^4}}\,
\cosh\left(2\sqrt{\mu}\overline{\tau}\right)\right\}\,,\end{equation}
for large enough values of the charge, namely, 
\begin{equation}q^4>\frac{\mu\ell^2(\Delta\mu)^2}{4}\,.\end{equation}
This solution is characterized by a minimum value of the scale factor
\begin{equation}
R_{min}^2=\frac{q^2}{2\mu}\left\{1+\sqrt{1-\frac{\mu\ell^2(\Delta\mu)^2}{4q^4}}
\right\}\end{equation}
obtained\footnote{The two times are related through
$\tau=\frac{q^2}{4\mu^{3/2}}\Bigl[2\sqrt{\mu}\,\overline{\tau}+
\left(1-\frac{\mu\ell^2(\Delta\mu)^2}{4 q^4}\right)^{1/2}\,
\sinh(2\sqrt{\mu}\,\overline{\tau})\Bigr]$.}
at $\overline{\tau}=\tau=0$ where the bouncing occurs.

The above lower limit on the charge should be compared with the upper limit
required by the non-extremality of the background\footnote{Since the curvature
term is always subdominant in the regime where the bounce occurs, for
simplicity we consider the $k=0$ condition.}, namely, 
\begin{equation}
\overline{q}^2<\overline{q}_+^2=\frac{2}{3\sqrt{3}}
\,min\left\{\overline{\mu}_{\pm}^3\right\}\,.\end{equation}
We are, thus, eventually led to the condition
\begin{equation}
(\mu_+-\mu_-)^2<\frac{32}{27}\frac{ min\left\{\mu_{\pm}^3\right\}}
{(\mu_++\mu_-)}\,.\end{equation}
Therefore, a bounce occurs with the black hole background possessing two
distinct horizons provided the asymmetry is not too large.

What about a radiation-dominated Brane? In that case, the Friedmann Equation
is of the form
\begin{equation}\dot{R}^2+k=\frac{1}{R^2}\left(\mu+\frac{2}{\ell^2}
\left(\frac{\hat{\rho}}{\sigma}\right)\right)
-\frac{q^2}{R^4}+\frac{1}{\ell^2}\left(\frac{\hat{\rho}}{\sigma}\right)^2
\frac{1}{R^6}+ \frac{\ell^2(\Delta\mu)^2}{16\left(\frac{\hat{\rho}}{\sigma}
\right)^2}\frac{R^2}{\left(1+\left(\frac{\sigma}{\hat{\rho}}\right)R^4\right)^2}\,,
\end{equation}
where we have introduced $\rho=\hat{\rho}R^{-4}$. For small values of the scale
factor, we can approximate this equation with
\begin{equation}\left(\frac{\dot{R}}{R}\right)^2\simeq \frac{1}{R^2}
\left(\mu+\frac{2}{\ell^2}\left(\frac{\hat{\rho}}{\sigma}\right)\right)
-\frac{q^2}{R^4}+\frac{1}{\ell^2}\left(\frac{\hat{\rho}}{\sigma}\right)^2
\frac{1}{R^6}\,.\end{equation}
Note that the asymmetry, in contrast to the empty-Brane case, contributes only
with a sub-leading term 
$\Bigl(\ell^2(\Delta\mu)^2\sigma^2/16\hat{\rho}^2\Bigr)R^2$, which 
can be dropped to a first approximation in our considerations concerning the
occurrence of a bounce. The remaining equation is identical to the one in the
symmetric case and yields essentially the same {\textit{lethal}} condition
\begin{equation}
1+\frac{1}{2}\,(\lambda_++\lambda_-)<\frac{4}{27}\,min\left\{\lambda_{\pm}^3
\right\}\,,
\end{equation}
with the $\lambda$'s being defined as 
$\mu_{\pm}=\lambda_{\pm}(8\pi G_4\hat{\rho}/3)$. The quantity
$(\lambda_++\lambda_-)/2$ is still constrained by nucleosynthesis to be
smaller than 0.14, a result which is in contradiction with the above
inequality: setting $(\lambda_++\lambda_-)/2 \simeq 0.13$, we are led
to the constraint $min\left\{\lambda_{\pm}\right\} > 1.9$, which cannot
be true given the constraint on their sum and the positive-definiteness
of $\lambda_\pm$.

In the case, finally, of a matter-dominated universe with 
$\rho=\tilde\rho/R^3$, the Friedmann equation (\ref{asym}) becomes
\begin{equation}
\dot{R}^2+k=\frac{\mu}{R^2} + 
\Bigl(\frac{\tilde{\rho}^2}{\ell^2 \sigma^2}-q^2\Bigr)\frac{1}{R^4}
+ \frac{2\tilde{\rho}}{\ell^2\sigma}\,\frac{1}{R}
+\frac{\ell^2(\Delta\mu)^2}{16\,(R^3 +\tilde{\rho}/\sigma)^2}\,.
\end{equation}
At large scales, the $(\Delta\mu)^2$-term has an $R^{-6}$ dependence
which makes this term negligible compared to the remaining ones. In 
the same way, at small scales, this term has the same scaling as the
curvature term and is again subdominant. Therefore, an asymmetric
bulk has no effect on the evolution of a matter-dominated universe.

\section{Effects of an Intrinsic Curvature Term}

It has been pointed out \cite{BAL} that the divergence arising for the
energy-momentum tensor at the boundary of the Schwarzschild-AdS space 
requires the introduction of an {\textit{intrinsic curvature}} scalar
counterterm. Such a term, arising in other frameworks as well \cite{DEF},
is certainly not forbidden. We shall, thus, assume the presence in the
action of the term \cite{CH}
\begin{equation}
\Delta {\cal{S}}=\frac{\beta \ell}{32\pi G}\int d^4x\,\sqrt{-\gamma}\,
{\cal R}_4\,,
\end{equation} 
where $\beta$ is a dimensionless constant that controls the ``turning on"
and ``off" of the boundary curvature term.
A priori, such an addition is mostly expected to modify the ``late", or
large-scale, evolution of the Brane and not the small-scale behaviour
responsible for the bounce. The resulting evolution equation on the
Brane, for a ${\cal Z}_2$-symmetric bulk spacetime, is
\begin{equation}2\sqrt{\dot{R}^2+f(R)}=\frac{8\pi G}{3}\,R\,(\rho+\sigma)-
\frac{\beta\ell}{2R}\left(\dot{R}^2+k\right)\,,\end{equation}
with our standard metric function 
$f(R)=k-\frac{\mu}{R^2}+\frac{q^2}{R^4}+\frac{R^2}{\ell^2}$. 
Taking the square of the above equation, we obtain a quadratic algebraic
equation with solution
\begin{eqnarray}
&~& \hspace*{-1.0cm} \dot{R}^2+k=\frac{8R^2}{(\beta\ell)^2}\left\{1+
\frac{\beta\ell}{2} \left(\frac{4\pi G}{3}\right)(\rho+\sigma)
\right.\nonumber \\[3mm] &~& \hspace*{2.5cm} \left.
-\sqrt{1+\frac{\beta^2}{4}+\beta\ell \left(\frac{4\pi G}{3}\right)(\rho+\sigma)
-\frac{(\beta\ell)^2}{4}\left(\frac{\mu}{R^4}-\frac{q^2}{R^6}\right)}
\right\}. \end{eqnarray}
In what follows, we will perform the same fine-tuning that, in the case 
$\rho=0$ and $\beta=0$, leads to a critical Brane, namely 
$\ell^{-1}=(4\pi G/3)\,\sigma$. 

Considering first the case of an empty Brane ($\rho=0$), we obtain the equation
\begin{equation}\dot{R}^2+k=\frac{4(\beta+2)}{(\beta\ell)^2}\,R^2\left\{1-
\sqrt{1-\left(\frac{\beta\ell}{\beta+2}\right)^2\left(\frac{\mu}{R^4}-\frac{q^2}{R^6}
\right)}\right\}\,.
\label{exact} \end{equation}
For small values of the parameter $\beta$, this equation can be replaced with
\begin{equation}\dot{R}^2+k \simeq \frac{2 R^2}{(\beta+2)}\left(\frac{\mu}{R^4}-
\frac{q^2}{R^6}\right)\,,\label{small-b}\end{equation}
which has the solutions displayed in Table 1 of section 4, with the parameter
rescaling
\begin{equation}
\mu\rightarrow \mu/\left(1+\frac{\beta}{2}\right)\,,  \qquad 
q^2\rightarrow q^2/\left(1+\frac{\beta}{2}\right)\,, \qquad
\epsilon^2\rightarrow \left(1+\frac{\beta}{2}\right)\epsilon^2\,.
\label{rescaled}\end{equation}
The smallness of $\beta$ required for the validity of the above
approximation is $\beta^2<<\mu^3/\ell^2q^4\sim q_+^2/q^2<1$.

For an appreciable value of $\beta$, such as the counterterm value $\beta=1$,
the above equation cannot be integrated analytically but the expectation that 
the small-scale behaviour is not going to be modified can be clarified by
some supportive arguments. Ignoring the curvature term proportional to $k$,
we can rewrite the expression under the square-root symbol in a manifestly
positive fashion in terms of the metric function $f(R)$, which is positive
for all points outside the outer horizon. Our equation is
\begin{equation}
H^2=\frac{4(\beta+2)}{(\beta\ell)^2}\left\{1-\sqrt{\frac{4(\beta+1)}{(\beta+2)^2}+
\frac{f(R)}{R^2}\left(\frac{\beta\ell}{\beta+2}\right)^2}\right\}\,.\end{equation}
Positivity of $H^2$ demands
\begin{equation}1>\frac{4(\beta+1)}{(\beta+2)^2}+
\frac{f(R)}{R^2}\left(\frac{\beta\ell}{\beta+2}\right)^2\,,\end{equation}
which turns out to be $\beta$-independent, namely
\begin{equation}\frac{f(R)}{R^2}<\frac{1}{\ell^2} \Longrightarrow 
R^2>\frac{q^2}{\mu}\equiv R_{min}^2\,.\end{equation}
Thus, the minimal value of the scale factor, obtained in this way, turns out to
be $\beta$-independent.

In the case of non-zero energy-density on the Brane ($\rho\neq 0$), our
Friedmann equation is
\begin{equation}H^2+\frac{k}{R^2}=\frac{8}{(\beta\ell)^2}\left\{1+
\frac{\beta}{2}\left(1+\frac{\rho}{\sigma}\right)-
\sqrt{\left(1+\frac{\beta}{2}\right)^2+\beta\left(\frac{\rho}{\sigma}\right)-
\frac{(\beta\ell)^2}{4}\left(\frac{\mu}{R^4}-\frac{q^2}{R^6}\right)}\right\}\,.
\end{equation}

Ignoring the $k$-term, we can repeat the argument we used in the $\rho=0$ case
and arrive again at a $\beta$-independent condition 
\begin{equation}\frac{\mu}{R^4}-\frac{q^2}{R^6}+\frac{1}{\ell^2}
\left[2\,\frac{\rho}{\sigma}+\left(\frac{\rho}{\sigma}\right)^2\right]>0\,,
\end{equation}
which, for example, for $\rho$ corresponding to radiation 
($\rho=\hat{\rho}R^{-4}$), leads to our well-known constraint 
$q^4>4(\hat{\rho}/\sigma \ell)^2(\mu+2\hat{\rho}/\sigma\ell)$.

At large scales, the curvature is, of course, expected to influence the
evolution. For large $R$, but not necessarily small $\beta$, we derive again 
Eq. (\ref{small-b}) for an empty Brane.  The solutions are again obtained from
Table 1 of section 4 by using the rescaled parameters (\ref{rescaled}). For the
cases $k=0, -1$, for which $R$ is eternally expanding after the bounce, there
is always a value of $R$ large enough for the approximation to be trusted for
any value of $\beta$. In the cyclic universe, however, obtained for $k=1$,
there is a maximum value of the scale factor given by $R_{max}^2=\mu/(\beta+2)$.
When substituted in the expression under the square-root in the exact equation
(\ref{exact}), with the charge term having been neglected as subdominant, 
a term $(\beta\ell)^2/\mu$ arises, which needs to be small compared to unity 
for our approximation to be valid. Even for values of $\beta$ of ${\cal O}(1)$,
this term is indeed negligible provided that the black hole mass-length parameter
$\sqrt{\mu}$ is much larger than the AdS length $\ell$. In that case, the
cyclic behaviour of the $k=1$ solution is retained for any value of $\beta$;
in the opposite case, only small values of $\beta$ are allowed.

\section{Conclusions and Discussion}

As in the case of an AdS bulk spacetime, the generalized Friedmann equation
derived on a Brane embedded in an AdS black-hole bulk spacetime allows for
modifications in the evolution of the four-dimensional subspace at small scales.
This result allows us to study the early time regime, as well as the late-time
regime for closed universes, and investigate whether the corresponding
cosmological singularities can be indeed avoided. The main attractive feature
of the Brane-World model considered in the present article, in which the 
five-dimensional spacetime is described by an AdS charged black hole, is the
fact that it realizes the bounce idea: the existence of a non-zero minimum
value of the scale factor that smoothly connects a contracting with an expanding
phase in the evolution of the four-dimensional subspace. In all cases considered,
this is indeed possible for a non-vanishing value of the charge parameter of the
five-dimensional black hole. The bounce effect therefore predicted in the
charged AdS black hole background provides support for a singularity-free
cosmology in which the Big Bang singularity is not present, as well as for a
{\textit{cyclic universe}}~\cite{CYCLIC} scenario in which neither the Big Bang
nor a Big Crunch singularity is present. 

Unfortunately, it is not possible to formulate a model, that would allow us to
study both the early and late time regimes in the history of the universe, since
different epochs are dominated by different energy-densities. It is therefore
necessary to distinguish between regions with smoothly connected but differing
equations of state, an approach followed here in chronological order. By
studying first, in Section 4, the case of an empty Brane with
either a zero or non-zero cosmological constant, embedded in a charged AdS
black-hole bulk spacetime, and joining together the two sets of solutions, we
were able to model a singularity-free, early inflationary era: the solutions 
are free from the Big Bang singularity and they smoothly interpolate to a
de Sitter expanding phase. The derived
constraints on the various parameters of the model put an upper bound on the
Hubble parameter of the asymptotic de Sitter phase which, when combined with
the demand that the magnitude of the density perturbations produced in this
period have the correct  size, lead to an intermediate-scale higher-dimensional
gravitational theory, i.e. $M_5 > 10^{-5} M_P$. 

We might assume instead that the early time regime is dominated by a
radiation-type equation of state. The derivation of the exact solution for the
scale factor on the Brane, at small scales, confirms the existence of a bouncing
and the absence of the Big Bang singularity. Assuming that this singularity-free,
radiation-dominated epoch lasts until the time of nucleosynthesis without
interruption, we are forced to satisfy a stringent constraint on the maximum
value of the radiation-type energy-density term that appears in the Friedmann
equation. As our analysis revealed, the range of parameters of the background,
for which the bouncing is possible and the nucleosynthesis constraint is
satisfied, exceeds the limit allowed by a non-extremal black hole and may
lead to an unstable background. This problem may be avoided by assuming that
the dominant equation of state does not remain the same for the whole range
of values from the bouncing point to the time of nucleosynthesis. Since the
universe must be radiation-dominated at nucleosynthesis time, that leaves two
options: (i) either the equation of state is dominated, at the bouncing point,
by the cosmological constant, which then leads to an inflationary period and
finally to a late radiation-dominated period, or (ii) an {\it early}
radiation-type equation of state gives way to an intermediate inflationary
period, as mentioned in Section 5, before coming back to a {\it late}
radiation-dominated period at the time of nucleosynthesis. 

As the universe expands, the radiation-dominated energy-density becomes
subdominant and gives its place to the matter-dominated one. In Section 6,
we studied the modifications that the generalized Friedmann equation brings
to the evolution of the Brane at this large-scale regime. In the case of
an open or flat Brane, the charge-dependent term is always negligible and
it is only the black-hole-mass-dependent term that survives and affects
the expansion rate of the Brane while preserving the eternal expansion
predicted by the four-dimensional Cosmological Model. In the case of a
closed Brane, the latter term causes an increase in the proper volume
of the universe but it cannot prevent the subsequent collapse. Assuming
that the equation of state remains matter-dominated during this late
small-scale regime, the charge-term becomes dominant and ensures the passage
from the contracting to a subsequent expanding phase, and thus the avoidance
of the Big Crunch. 

However, cosmological observations~\cite{OBS}\cite{OBS1}\cite{OBS1-1}\cite{OBS2} 
strongly indicate that the present universe is spatially flat and accelerating
due to some dominant dark-energy component. The simplest possibility is that this dark
energy of unknown origin is in the form of a small cosmological constant that
puts the universe in an indefinitely expanding de Sitter phase. This scenario
can be easily accommodated in the framework of the second set of 
solutions derived in Section 4, that predict an asymptotic de Sitter expansion
for all values of $k$. The same solutions could also model the alternative
scenario in which the dark energy is generated by a slowly varying scalar
field~\cite{Q}, with a $w\simeq-1$ equation of state and thus an almost constant
energy-density. In such a scenario, the derived de Sitter expanding phase is
only an intermediate one that eventually will give way to an asymptotic Minkowski
regime as the speed of expansion will start decreasing. In both cases, it is
only the black-hole mass parameter that is relevant to the present-time
evolution, by restricting the Hubble parameter for the, either eternal or
temporary, de Sitter expansion phase, while the charge parameter has
absolutely no effect. 

In addition to the uncertainty about the presently valid equation of state, 
the very late evolution is also open to speculation and conjecture leaving open
the possibility of a contracting and, perhaps, cyclic, behaviour. If, for
example, the dark-energy eventually becomes negative, the universe will
collapse~\cite{LINDE}. In the cosmic contraction scenario, the background
charge will be essential in avoiding a Big Crunch and bouncing back into
an expanding state, just like in the case of a matter-dominated phase.
Nevertheless, the specific energy-density required for a late-time contracting
phase has to be inserted in the Brane energy-momentum tensor in an arbitrary
fashion. The fundamental physics associated with its required form is still
lacking. An interesting, and perhaps fruitful approach, would be to try to
investigate ways of non-trivial Bulk-Brane interactions resulting, through
the exchange of energy~\cite{TOM}, in a dynamical evolution of the equation
of state on the Brane that accounts for the present time accelerating phase
as well as for a possible contracting one.

Let us finally note that, according to our analysis conducted in Sections
7 and 8, variants of the above model, in which the bulk spacetime is assumed to
be asymmetric or a Brane curvature term is added in the
action, do not lead to any radical changes in the type of behaviour encountered
near, or  the existence itself of, the bouncing point. In the first variant,
it is only in the case of an empty Brane and for large values of the 
black-hole mass difference on the two sides of the Brane, that
the extra term in the Friedmann equation tends to prohibit the occurrence
of the bouncing. In every other case, the effect of this term is irrelevant.
In the second variant, the Brane curvature term has an effect only at large
scales, as expected, and can be ignored at the time of the bouncing, either
at early or late times, without any loss of information.

\bigskip
{\textbf{Acknowledgments}}

P.K. would like to thank the Theoretical Physics Group at the Technical
University of Munich, where parts of this work were completed, for its
hospitality and financial support.
K.T. would like to thank the CERN Theory Division for its hospitality and
A. Petkou for useful discussions. He also acknowledges traveling support
from the RTN programme HPRN-CT-2000-00152.

\end{document}